\documentclass[times, 10pt,twocolumn]{article} 
\usepackage{latex8}
\usepackage{times}
\usepackage{amsmath,graphicx}
\usepackage{amssymb}
\usepackage{xcolor}

\begin{document}

\title{Attention-Guided Generative Adversarial Network to Address Atypical Anatomy in Modality Transfer}

\author{Hajar Emami\\
Wayne State University\\ Department of Computer Science \\ Detroit, MI 48202, USA\\ hajar.emami.gohari@wayne.edu\\
\and
Ming Dong\\
Wayne State University\\ Department of Computer Science \\ Detroit, MI 48202, USA\\ mdong@wayne.edu\\
\and
Carri K. Glide-Hurst\\
Henry Ford Health System\\Department of Radiation Oncology\\Detroit, MI 48202, USA\\churst2@hfhs.org\\
}

\providecommand{\Keywords}[1]{\textbf{\textit{Keywords---}}  #1}

\maketitle
\thispagestyle{empty}

\begin{abstract}
Recently, interest in MR-only treatment planning using synthetic CTs (synCTs) has grown rapidly in radiation therapy. 
However, developing class solutions for medical images that contain atypical anatomy remains a major limitation. 
In this paper, we propose a novel spatial attention-guided generative adversarial network (attention-GAN) model to generate accurate synCTs using T1-weighted MRI images as the input to address atypical anatomy.
Experimental results on fifteen brain cancer patients show that attention-GAN outperformed existing synCT models and achieved an average MAE of 85.22$\pm$12.08, 232.41$\pm$60.86, 246.38$\pm$42.67 Hounsfield units between synCT and CT-SIM across the entire head, bone and air regions, respectively. Qualitative analysis shows that attention-GAN has the ability to use spatially focused areas to better handle outliers, areas with complex anatomy or post-surgical regions, and thus offer strong potential for supporting near real-time MR-only treatment planning.
\end{abstract}

%

\vspace{0.5cm}
\Section{Introduction}

Computed tomography (CT) imaging has been widely used as the primary image modality for dose calculation in radiation therapy planning (RTP). Since magnetic resonance imaging (MRI) has superior soft tissue contrast that improves target and organ at risk (OAR) segmentation accuracy in the brain, it is currently used as an adjunct to CT in radiation therapy. The current standard of care for RTP is based on CT simulation (CT-SIM) with registration to corresponding MRI datasets \cite{pietrzyk1994interactive}. Recent interest has been given to MR-only radiation therapy treatment planning in which synthetic CT is generated from MRI data. 
MRI-only treatment planning eliminates co-registration geometrical uncertainties, reduces costs and patient time commitment, and eliminates the radiation burden of an additional CT scan. 

Many techniques have been proposed in the literature for generating synCTs from MRI data. Atlas-based methods typically consist of performing a deformable registration between a previously developed atlas (or multiple atlases) and a test MR image, and estimating an attenuation map via image warping \cite{kops2007alternative}. Atlas-based methods often rely on deformable image registration between datasets that may introduce additional uncertainty while also being difficult to validate. Learning-based method directly learn a non-linear mapping between CT and MRI images. Johansson et al. \cite{johansson2011ct} proposed to train a Gaussian mixture regression model to generate synCTs from ultra-short echo time (UTE) and T2-weighted MR images. Recently, deep learning methods have been proposed for synCT generation, showing superior performance to the atlas-based and conventional machine learning approaches. In \cite{han2017mr} and \cite{dinkla2018mr}, convolutional neural network (CNN) models were developed to perform image-to-image mapping between MR and CT brain datasets. Nie et al. \cite{nie2016estimating} utilized a 3D fully convolutional network (FCN) to generate synthetic CT images from MRI data. Generative adversarial network (GAN)\cite{goodfellow2014generative} models with two competing networks were developed to generate synCTs from MRI data \cite{nie2017medical, emami2018generating} and achieved superior results over FCN models. In \cite{wolterink2017deep}, a CycleGAN \cite{zhu2017unpaired} model consisting of two synthesis generators and two discriminators was employed on unpaired MRI and CT data and used to transform MR images into CT images and vice versa.

All aforementioned techniques including deep learning based models have difficulties with generating atypical anatomy, outliers or post-surgical regions. In MR-only treatment planning, it is critical to ensure regions such as air, bone and tissues are accurately generated in synCT, to yield high geometric fidelity and ensuring high accuracy in dose calculation as compared to the original CT-based treatment plan. Inspired from human attention mechanism \cite{rensink2000dynamic}, attention-based models have gained popularity in a variety of computer vision and machine learning tasks including image classification \cite{xiao2015application, zhou2016learning}, image segmentation \cite{chen2016attention}, image-to-image translation \cite{mejjati2018unsupervised, emami2020spa,emami2020frea}, natural language processing \cite{eriguchi2016tree, lin2017structured} and time series forecasting \cite{siridhipakul2019multi, aliabadi2020attention}. Attention improves the performance by encouraging the model to focus on the most important regions.
Building upon our previous work \cite{emami2018generating}, we propose a novel spatial attention-guided generative adversarial network (attention-GAN) to minimize the spatial difference in these regions in synCTs that can better handle atypical anatomies and outliers. Attention-GAN computes the spatial attention in its discriminator and draws attention of the generator to regions that cause the most difference between the real CT and synCT images.
Experimental results show that attention-GAN outperforms existing synCT models, offering strong potential for supporting MR-only radiation therapy workflows.

\begin{figure*}[!htb]
\begin{center}
  \centering
  \centerline{\includegraphics{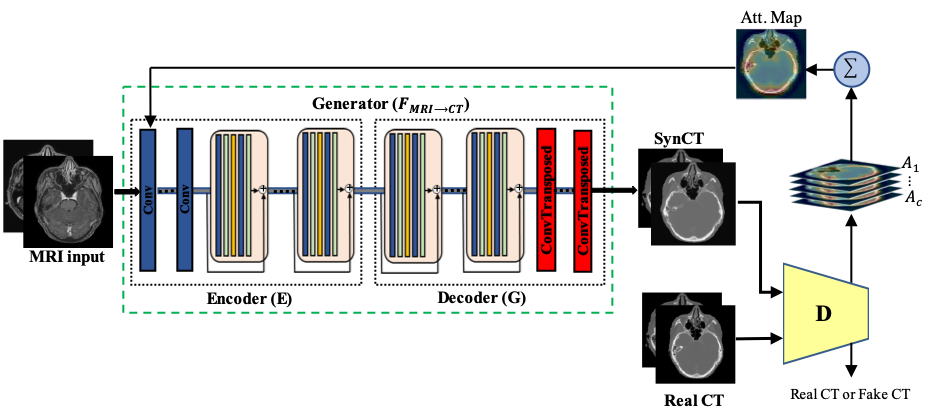}}
  \vspace{-0.4cm}
  \caption{The architecture of the proposed attention-GAN.}
   \vspace{-1cm}
\label{fig:architecture}
\end{center}
\end{figure*}

\vspace{0.3cm}
\Section{Materials and Method}
Recently, attention mechanism was introduced in image-to-image translation tasks to decompose the generative network into two separate networks: the attention network to predict regions of interest and the transformation network to transform the image from one domain to another. Then, the attention is employed to separate foreground and background in an input image \cite{mejjati2018unsupervised,chen2018attention}. Different from prior attention work \cite{mejjati2018unsupervised,chen2018attention} that use an additional network to generate attention maps, our spatial attention is defined as the spatial map showing the areas that the discriminator focuses on when classifying an input image as real or fake. The extracted spatial attention is fed back to the generator to help the generator detect and focus on regions that cause the most difference between the real CT and synCT images.

\vspace{0.5cm}
\SubSection{Network Architecture}
The architecture of the proposed Attention-GAN model is shown in Fig. 1. The generator $F_{MRI \rightarrow CT}$ is an encoder-decoder with three down-sampling convolutional layers and four residual blocks \cite{he2016deep} in the encoder $E$. The decoder part $G$ includes four residual blocks followed by three transposed convolutional layers. All residual blocks have shortcut connections (shown as solid arrows) used for adding the input of each block to its output to skip one or more layers. These shortcut connections can ease the training of the generator without adding extra parameters or computational complexity. The discriminator is a regular CNN with six convolutional layers followed by batch normalization and ReLU that classifies an input image as real or synthetic CT.

\vspace{0.5cm}
\SubSection{Spatial Attention}
Incorporating spatial attention into GAN helps the generator focus on the regions of interest like bone, air, and interfaces to produce synCTs with better agreement with corresponding CT dataset. In synCT GAN, the discriminator classifies the input to either synCT or real CT. Attention is computed as the sum of the absolute values in different layers of the discriminator across the channel dimension:
\begin{align}
W_{att}(A)&= \sum\limits_{i=1}^C |A_i|
\end{align}
where $A_i$ is i-th feature map of a discriminator layer and C is the number of channels. The computed attention is a good indicator of the importance of neuron activations with respect to the given input in that specific layer \cite{zagoruyko2016paying}.

The focus of attention maps varies at different layers in the discriminator. For instance, when first classifying a CT image, the first layer's activation level is high for low-level features whereas in the middle layers, the activation level is higher on the most discriminative regions in the brain such as air and bone regions. Later layers typically focus on the entire head in the image. In order to focus more on the most discriminative regions, the proposed attention-GAN model extracts the spatial attention maps from the middle layer of the discriminator, which is correlated to discriminative regions in brain images. These attention maps of CT/synCT illustrate the areas where the discriminator focuses in order to correctly classify the CT image as real or fake.

The extracted spatial attentions are normalized to the range of $[0,1]$, up-sampled and brought back into the first layer of the encoder. It is worth pointing out that earlier approaches on attention-guided GANs \cite{mejjati2018unsupervised,chen2018attention} require loading generators, discriminators and additional attention networks into the GPU memory all at once, which may cause computational and memory limitations. In comparison, our attention-GAN model is a lightweight model that does not need additional attention networks, attention maps are directly obtained from the discrimanator.

\SubSection{Loss Function}
The proposed attention-GAN model learns a mapping from input MR images to CT images. The discriminator loss $\cal{L}_D$ and the generator loss $\cal{L}_F$ for the mapping $F: MRI \rightarrow CT$ are given as:
\begin{align}
{\cal{L}}_{D}&= \frac{1}{2}\mathbb{E}_{ct{\sim}Pdata(ct)}[(D(ct)-1)^2]\nonumber\\
                              &+\frac{1}{2} \mathbb{E}_{mri{\sim}Pdata(mri)}[D(G(E(W_{att}*mri)))^2]
\end{align}
and
\begin{small}
\begin{align}
 {\cal{L}}_{F}&= \frac{1}{2}\mathbb{E}_{mri{\sim}Pdata(mri)}[(D(G(E(W_{att}*mri)))-1)^2]+\lambda{\cal{L}}_{L_1}
\end{align}
 \end{small}
where * is an element-wise product and the attention weight $W_{att}$ computed in Eq. 1 is applied to the feature map in the first layer of the encoder $mri$. We replace the negative log likelihood objective by a least squares loss which has been shown to be more stable during training and generates higher quality results \cite{mao2017least}. Prior research on image to image translation shows that adding $L_1$ distance to the GAN objective as a part of the reconstruction error can also help generate more realistic output images \cite{isola2017image}:
\begin{align}
{\cal{L}}_{L_1}&= \mathbb{E}_{mri,ct{\sim}Pdata(mri,ct)}[||ct-G(E(W_{att}*mri))||_1]
\end{align}

Since the goal of the generator is to minimize the loss function and adversarially the goal of the discriminator is to maximize it, the final objective can be expressed as:
\begin{align}
F^*,D^* &=\arg \, \smash{\displaystyle\min_{F}} \; \smash{\displaystyle\max_{D}} {\cal{L}}(F,D)+\lambda{\cal{L}}_{L_1}
\end{align}
where we set the loss hyper-parameter $\lambda$ = 10 throughout our experiments.
In the proposed model, the generator can focus more on the most discriminative regions based on the attention maps received from the discriminator, leading to more accurate synCTs.

\vspace{0.5cm}
\SubSection{Training Dataset and Preprocessing}
Fifteen brain cancer patients that had undergone both CT simulation (CT-SIM) and MR simulation (MR-SIM) were retrospectively analyzed as part of an Institutional Review Board approved study. MR-SIM was performed on a 1.0T Panorama High Field Open (Philips Medical Systems, Best, Netherlands). In this study, post-Gadolinium T1-weighted images were acquired for each patient using an 8-channel head coil with a voxel size of 0.90$\times$0.90$\times$1.25 $mm^3$. All brain CT images were acquired using a Brilliance Big Bore (Philips Health Care, Cleveland, OH) scanner at the following settings: 500 mAs, 512$\times$512 in-plane image dimensions, 0.88$\times$0.88 $mm^2$ in-plane spatial resolution, and 1-mm slice thickness.

All MR images are then aligned to their corresponding CTs using the intra-subject rigid registration, in which the size and spacing of all paired CT and their corresponding MR images are matched. All voxels outside of the patient external contour including the immobilization devices and couch structures were removed in all CT slices to have more precise evaluation of the generated synCTs. A binary head mask was derived from each MR slice using thresholding and morphological operators to separate the head region from the immobilization device. No inter-subject registration and contrast adjusting are used in this work. 

\vspace{0.3cm}
\SubSection{Comparison with Other SynCT Models}
In order to compare the proposed attention-GAN model with existing synCT models, we implemented and trained both a CNN with no discriminator block and a regular GAN and evaluated them on the same brain dataset. In order to have a fair comparison, we use the same architecture of the proposed attention-GAN's generator in both CNN and the generator in the regular GAN but without spatial attention. All models are trained with the same training parameters for the same number of epochs.

\vspace{0.5cm}
\SubSection{Quantitative Measurements}
To quantitatively evaluate the prediction accuracy of the model between synCT and CT-SIM data, three well-established metrics were used: mean absolute error (MAE), peak signal-to-noise ratio (PSNR) and structural similarity index (SSIM).
\begin{align}
MAE&=\frac{\sum\limits_{i=1}^H |realCT(i)-synCT(i)|}{H}
\end{align}
 \begin{align}
PSNR&=10log_{10}(\frac{Q^2}{MSE})
\end{align}
\begin{small}
 \begin{align}
SSIM&=\frac{(2\mu_{realCT}\mu_{synCT} + C_1)(2\delta_{realCTsynCT}+C_2)}{(\mu_{realCT}^2+\mu_{synCT}^2+C_1)(\delta_{realCT}^2+\delta_{synCT}^2+C_2)}
\end{align}
\end{small}
In Eq. (6), H is the number of head voxels. The parameters $C_1=(K_1Q)^2$ and  $C_2=(K_2Q)^2$ are two variables to stabilize the division, where $K_1=0.01$ and $K_2=0.02$. In Eq. (7), $Q$ is the maximum intensity value of realCT and synCT images, and $MSE$ is the mean squared error. The lower the MAE, the higher the accuracy of the synCT images, that is, better agreement with the real CTs. For both SSIM and PSNR, higher values give better prediction.

\begin{table}
\centering
\caption{Mean absolute error (MAE), peak signal-to-noise ratio (PSNR) and structural similarity index (SSIM) computed between real CTs and synCTs for five-fold cross validation. MAEs are in Hounsfield units (HU).}
\label{table:Table1}
\begin{tabular}{c|cccc}
\hline
   Fold \# & MAE(HU) & PSNR &	SSIM \\
\hline
1 & 83.59 & 28.30 & 0.85   \\
2 &  80.12 & 29.43 & 0.85 \\
3 & 89.90 & 27.16 & 0.83 \\
4 & 85.36 & 27.75  & 0.84  \\
5 & 88.47 & 27.39  & 0.83 \\
\hline
\end{tabular}
\end{table}

\begin{table}
\centering
\caption{Average MAE computed from full field of view (full FOV), air, bone, and tissue regions for 15 cases.}
\label{table:Table1}
\begin{tabular}{c|ccccc}
\hline
     & FOV & Air & Bone  & Tissue\\
   Patient & MAE & MAE & MAE & MAE\\
                & (HU) & (HU) & (HU) & (HU) & \\
\hline
Mean & 85.22 & 232.41 & 246.38 & 39.13\\
SD & 12.08 & 60.86 & 42.67 & 7.89\\
\hline
\end{tabular}
\end{table}

\begin{figure*}[htb]
\begin{center}
\centering
   \includegraphics[width=1\linewidth]{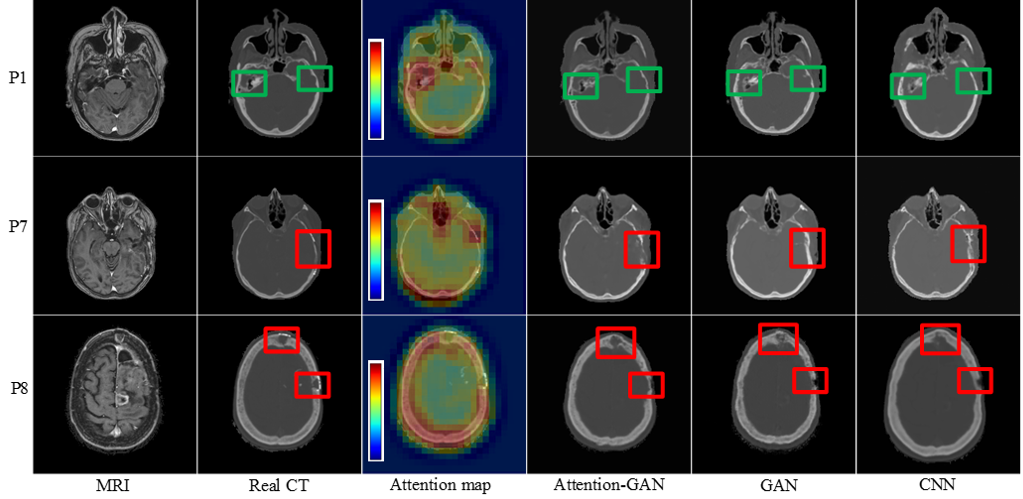}\\
\vspace{-0.7cm}
\hfill
\caption{Qualitative comparison of synCTs generated with attention-GAN, regular GAN and CNN for three different patients. Attention maps have values in range \textbf{$[0,1]$}.}
   \vspace{-0.9cm}
\label{fig:results}
\end{center}
\end{figure*}

\vspace{0.5cm}
\Section{Experimental Results}

\SubSection{Experimental Setup}
   \vspace{-0.1cm}
The attention-GAN generator and discriminator networks are trained on paired MRI-CT images. The initial weights are from a Gaussian distribution with parameter values of 0 and 0.02 for mean and standard deviation. The model is trained with ADAM optimizer \cite{kingma2014adam} with an initial learning rate of 0.0002 and with a batch size of 1. The  model stabilized after 200 epochs with no significant change in PSNR, SSIM and MAE values.

\vspace{0.5cm}
\SubSection{Attention-GAN Results}
   \vspace{-0.2cm}
To validate the performance of the proposed attention-GAN model, a five-fold cross validation was used in a similar manner to recent works \cite{han2017mr, emami2018generating}. That is, 15 cases are randomly partitioned into five groups, and in each experiments, four groups are selected for training and one group for testing phase. The attention-GAN training took approximately 11 hours with a GeForce 980 Ti GPU, and the testing takes about 6 seconds for generating synCT of a new MRI data. The MAE, PSNR and SSIM metrics computed based on the real and synCTs are reported for each fold in Table 1.

Table 2 summarizes the average of the three quantitative measurements on full field of view (FOV) of the entire head for all fifteen patients. The average MAE is 85.22$\pm$12.08 Hounsfield units (HU); The average PSNR is 28.02$\pm$1.14, and the average SSIM is 0.84$\pm$0.03. To better evaluate the proposed model, the average MAEs on different regions of the brain, including bone, air and soft tissue are computed and summarized in Table 2. Ground truth and synCT air masks were determined using a threshold of -465 HU, which was the average of optimal values for the upper airway \cite{nakano2013new}. For bone regions, a threshold of +200 HU was used in both real and synCTs \cite{emami2018generating,nakano2013new}. The average MAE observed in segmented bone and air regions are 246.38$\pm$42.67 HU and 232.41$\pm$60.86 HU, respectively. Lower MAE of 39.13$\pm$7.89 HU was achieved in the remaining tissue regions.

\vspace{0.5cm}
\SubSection{Comparison with Other SynCT Models}
   \vspace{-0.1cm}
The average MAE obtained in FOV for all fifteen patients using the CNN and GAN with no attention are 102.41$\pm$19.94 HU and 89.30$\pm$10.25 HU, respectively. Also, the average MAE observed in segmented bone and air regions using GAN with no attention is 255.22$\pm$47.74 and 240.94$\pm$60.21, respectively.

Fig. 2 highlights three randomly selected synCT patient cases generated by the proposed attention-GAN, regular GAN with no attention and CNN. Qualitative comparison for the three different methods show that the synCTs generated by attention-GAN are more similar to the real CTs, less noisy, and have more preserved details when compared to CNN and regular GAN results. The attention maps obtained from the discriminator (the third column in Fig. 2) show higher weights around bone and air regions, helping attention-GAN synthesize regions near the bone/air interfaces (highlighted in green boxes) more accurately. 
Finally, P7 and P8 (the second and third rows) represent patients who had post-surgical resection resulting in anomalies in the skull after surgery. Atypical anatomy presents great challenges for applying deep learning models in medical imaging. For these two patients, our results show that attention-GAN model well represents the bone that was affected by surgery (highlighted in red boxes) and generates more accurate synCTs when compared to other existing models without attention mechanism. These qualitative results show that attention-GAN has the ability to use spatially focused areas to better handle outliers or bone/air interfaces, and anomalies in the patient due to surgery.

\Section{Conclusion}
   \vspace{-0.3cm}
In this paper, we introduced a novel attention-GAN model to generate synCT using MRI as the input. Experimental results show that attention-GAN achieves superior results on synCTs than current models, offering strong potential for supporting MR-only workflows. The development of this pipeline is an important step to apply the methodology to other disease sites such as generating synCTs for patients with prosthetic hips, implanted markers, or other atypical cases. Further development and validation will enable a flexible, widely applicable pipeline to support different patient conditions for MR-only treatment planning. In the future, we plan to adopt the spatial attention in other synCT generation settings including multi-modality and unpaired workflows, and also to extend to other disease sites.

\nocite{ex1,ex2}
\bibliographystyle{latex8}
\bibliography{latex8}

\end{document}